\documentclass[twocolumn]{autart}    

\usepackage{graphicx}          
\usepackage{amsmath} 
\usepackage{amssymb}  
\usepackage{algorithm}
\usepackage[noend]{algpseudocode}
\usepackage{color}
\usepackage{graphicx}
\usepackage{booktabs}

\DeclareMathOperator*{\argmin}{arg\,min}

\newcommand{\bu}{\mathbf{u}}

\newcommand{\x}{\mathbf{x}}

\newcommand{\y}{\mathbf{y}}

\begin{document}

\begin{frontmatter}

\title{Learning a Better Control Barrier Function \\Under Uncertain Dynamics\thanksref{footnoteinfo}}

\thanks[footnoteinfo]{An earlier version~\cite{DBLP:journals/corr/abs-2205-05429} of a portion of this paper was presented at the 61st IEEE Conference on Decision and Control, Cancun, Mexico, December 2022.}

\author[NYU]{Bolun Dai}\ead{bd1555@nyu.edu}, 
\author[NYU]{Prashanth Krishnamurthy}\ead{prashanth.krishnamurthy@nyu.edu},
\author[NYU]{Farshad Khorrami}\ead{khorrami@nyu.edu}
\address[NYU]{Control/Robotics Research Laboratory (CRRL), Dept. of Electrical and Computer Engineering,\\ New York University Tandon School of Engineering, Brooklyn, NY 11201}
          

\begin{abstract}
Using control barrier functions (CBFs) as safety filters provides a computationally inexpensive yet effective method for constructing controllers in safety-critical applications. However, using CBFs requires the construction of a valid CBF, which is well known to be a challenging task, and accurate system dynamics, which are often unavailable. This paper presents a learning-based approach to learn a valid CBF and the system dynamics starting from a conservative handcrafted CBF (HCBF) and the nominal system dynamics. We devise new loss functions that better suit the CBF refinement pipeline and are able to produce well-behaved CBFs with the usage of distance functions. By adopting an episodic learning approach, our proposed method is able to learn the system dynamics while not requiring additional interactions with the environment. Additionally, we provide a theoretical analysis of the quality of the learned system dynamics. We show that our proposed learning approach can effectively learn a valid CBF and an estimation of the actual system dynamics. The effectiveness of our proposed method is empirically demonstrated through simulation studies on three systems, a double integrator, a unicycle, and a two-link arm.
\end{abstract}

\end{frontmatter}

\section{Introduction}
Ensuring safety is crucial when designing controllers for real-world applications~\cite{DaiKKGTK23}\cite{DaiKPK23}\cite{DaiHKK23}\cite{DaiKKK23}. With the increasing usage of automated systems, e.g., self-driving cars~\cite{DBLP:journals/corr/BojarskiTDFFGJM16}, the ability to guarantee the safety of such systems becomes increasingly important. In optimal control, safety is often ensured by casting the safety requirements as constraints~\cite{DBLP:conf/iros/HowellJM19}. However, when the optimization problem gets larger~\cite{DBLP:journals/siamrev/Kelly17}, the solution time increases rapidly, limiting its usage for guaranteeing safety in complex environments, which usually requires the control system to react quickly. Recently, Hamilton-Jacobi reachability analysis has been used to generate safe controls~\cite{DBLP:journals/corr/abs-1709-07523}\cite{mitchell2002application}. When solved offline, it provides a way to generate safe controls quickly online. However, its usage is greatly limited by the curse of dimensionality~\cite{DBLP:journals/corr/abs-1709-07523}. With the rise in popularity of learning-based methods in control synthesis, learning-based methods have also been used to synthesize controllers for safety-critical tasks~\cite{DBLP:journals/corr/abs-2107-07931}\cite{ThananjeyanBNLS21}\cite{BansalT21}. However, most learning-based methods require a significant amount of unsafe interactions to learn a safe controller~\cite{DBLP:journals/ral/QinSF22}, which might be costly or impossible to obtain.

Another popular method to synthesize safe control is utilizing control barrier functions (CBFs)~\cite{DBLP:conf/eucc/AmesCENST19}. CBFs can be used with control Lyapunov functions (CLFs) or as a safety filter for an unsafe performance controller~\cite{DBLP:conf/eucc/AmesCENST19}. In both cases, the control can be obtained by solving a quadratic program (QP)~\cite{DBLP:journals/tac/AmesXGT17} which can be done at a very high frequency using modern optimization solvers. Given its many advantages, CBFs have been used on many safety-critical tasks, e.g., biped and quadrupedal locomotion on stepping stones~\cite{DBLP:conf/cdc/NguyenHGAS16}~\cite{GrandiaTA021}, adaptive cruise control~\cite{DBLP:conf/cdc/AmesGT14}, and multi-agent aerial maneuver~\cite{DBLP:conf/iclr/QinZCCF21}.

Although CBF provides a promising direction in safe controller synthesis, there are two significant assumptions when applying CBF-based controllers: having access to a valid CBF and having accurate system dynamics. A common approach to finding a valid CBF is to start with a description of the safe set, usually in the form of state constraints, and find a function that is positive only within the safe set and has the appropriate relative degree with respect to the system dynamics. This method is plausible for simple constraints. However, finding a valid CBF that recovers the entire safe set becomes increasingly challenging~\cite{DBLP:conf/cdc/ChoiLSTH21} as the constraints become nonlinear or nonconvex. To mitigate this issue, work has been done in learning the CBF. In~\cite{DBLP:conf/iros/SaverianoL19}, human demonstrations have been used to map the boundaries of the safe set, and a CBF is then learned. This method may not scale to constraints in higher dimensions. Instead of having information on safe set boundaries, work has been done on utilizing expert demonstrations of safe and unsafe trajectories~\cite{DBLP:conf/cdc/RobeyHLZDTM20}. Additionally, work has been done in learning CBFs using data collected online. In~\cite{DBLP:journals/corr/abs-2106-05341}, a CBF is synthesized using only onboard sensors. 

The aforementioned learning-based methods assume no knowledge of the CBF and learn it from scratch. This is an overly restricting assumption because handcrafting a conservative CBF is usually possible in many cases. Recently, work has been done in learning a CBF starting from an initial conservative CBF. In~\cite{TonkensH22}, an HCBF is used to warm start a dynamic program that refines the HCBF to enlarge the recovered safe set. In~\cite{DBLP:journals/corr/abs-2205-05429}, a learning-based approach is used to learn the difference between a conservative HCBF and a CBF that recovers a more significant portion of the safe set. 

Another assumption made in many CBF-related works is having access to the system dynamics, which is usually not the case in real-world applications~\cite{DBLP:journals/corr/abs-2107-11836}. Work has been done in learning the CBF in a model-free fashion~\cite{DBLP:journals/ral/QinSF22}. However, like learning CBFs from scratch, having no knowledge of the system dynamics is also overly restricting since an approximate nominal model of the system dynamics is often known in many real-world applications. Recently, work has been done in learning the system dynamics for CBF-based controllers~\cite{TaylorSYA20}\cite{WangMLS021} while assuming access to a ground truth CBF. In this paper, we build on our earlier work in learning-based CBF refinement~\cite{DBLP:journals/corr/abs-2205-05429} and further develop and evaluate the methodology under uncertain system dynamics.


In this paper, we propose an algorithmic approach to learn both the CBF and the system dynamics starting from an HCBF and a nominal model of system dynamics. The main contribution of this paper is threefold: (1) starting from an HCBF, we develop a method to learn a well-behaved CBF that recovers a more significant portion of the safe set (also known as CBF refinement~\cite{TonkensH22}) using a CBF prior (i.e., distance function); (2) we extend the CBF refinement problem to include problems with uncertain dynamics; (3) we show the effectiveness of our proposed approach using extensive simulation studies on three systems: double integrator, unicycle, and a two-link arm. The remainder of this paper is structured as follows. In Section II, the foundations of CBF are briefly summarized. In Section III, the problem formulation is given. In Section IV, the proposed method is presented. In Section V, the results of the simulation studies on a double-integrator, a unicycle, and a two-link arm are presented. Section VI concludes the paper with a summary and discussion of future works.
\section{Preliminaries}
In this section, we review the concept of CBF and how it is utilized in safety-critical applications. Consider a control affine system
\begin{equation}
    \dot{\x} = \mathbf{f}(\x) + \mathbf{g}(\x)\bu,
    \label{eq:control_affine_system}
\end{equation}
where the state is represented as $\x\in\mathbb{R}^n$ and the control as $\bu\in\mathcal{U}\subset\mathbb{R}^m$, with $\mathcal{U}$ being the admissible set of controls. The locally Lipschitz continuous functions $\mathbf{f}: \mathbb{R}^n\rightarrow\mathbb{R}^n$ and $\mathbf{g}: \mathbb{R}^n\rightarrow\mathbb{R}^{n\times m}$ represent the drift and the control influence matrix, respectively. We assume access to a feedback controller
\begin{equation}
    \bu = \pi(\x),
    \label{eq:feedback_controller}
\end{equation}
with $\pi: \mathbb{R}^n\rightarrow\mathbb{R}^{m}$ also being a locally Lipschitz continuous function. Substituting~\eqref{eq:feedback_controller} into~\eqref{eq:control_affine_system}, the closed-loop dynamics are given by:
\begin{equation}
    \dot{\x} = \mathbf{f}_{\mathrm{cl}}(\x) = \mathbf{f}(\x) + \mathbf{g}(\x)\pi(\x).
    \label{eq:closed_loop_system}
\end{equation}
For any initial state $\x_0\in\mathbb{R}^n$, there exists a maximal time interval of existence
\begin{equation}
    \mathbf{I}(\x_0) = [t_0, t_{\mathrm{max}}),
\end{equation}
where $\x(t)$ is a unique solution to~\eqref{eq:closed_loop_system} on $\mathbf{I}(\x_0)$; when $t_{\mathrm{max}} = \infty$, the system defined in~\eqref{eq:closed_loop_system} is considered forward complete~\cite{Khalil15}.

The notion of safety is defined for this work as forward invariance with respect to the safe set $\mathcal{C}\subset\mathbb{R}^n$:

\begin{defn}[Forward Invariance \& Safety]
\label{def:forward_invariance_safety}
The system defined in~\eqref{eq:closed_loop_system} is forward invariant with respect to $\mathcal{C}$ if for every $\x_0\in\mathcal{C}$, we have $\x(t)\in\mathcal{C}$ for all $t\in\mathbf{I}(\x_0)$. A system that is forward invariant with respect to $\mathcal{C}$ is said to be safe with respect to $\mathcal{C}$. A controller that makes a closed-loop system safe with respect to $\mathcal{C}$ is said to be safe with respect to $\mathcal{C}$.
\end{defn}
We consider $\mathcal{C}$ to be the 0-superlevel set of a continuously differentiable function $h:\mathbb{R}^n\rightarrow\mathbb{R}$, yielding
\begin{subequations}
\label{eq:CBF_conditions}
\begin{align}
    \mathcal{C} &= \{\x\in\mathbb{R}^n \mid h(\x) \geq 0\},\\
    \partial\mathcal{C} &= \{\x\in\mathbb{R}^n \mid h(\x) = 0\},\\
    \mathrm{Int}(\mathcal{C}) &= \{\x\in\mathbb{R}^n \mid h(\x) > 0\},
\end{align}
\end{subequations}
where $\partial\mathcal{C}$ represents the boundary of $\mathcal{C}$ and $\mathrm{Int}(\mathcal{C})$ represents the interior of $\mathcal{C}$. Additionally, we assume that $\mathrm{Int}(\mathcal{C})$ is not an empty set, i.e., $\mathrm{Int}(\mathcal{C})\neq\emptyset$, and that $\mathcal{C}$ does not contain any isolated points. Before defining CBFs, we first define extended class $\mathcal{K}$ functions:

\begin{defn}[Extended class $\mathcal{K}$ function]
A continuous function $\alpha: (-b, a)\rightarrow\mathbb{R}$ is called an extended class $\mathcal{K}$ function when $\alpha(0) = 0$ and $\alpha$ is strictly monotonically increasing. When $a = \infty$, $b = \infty$, and
\begin{equation*}
    \lim_{r\rightarrow\infty}\alpha(r) = \infty\ \ \ \ \mathrm{\&}\ \ \ \ \lim_{r\rightarrow-\infty}\alpha(r) = -\infty,
\end{equation*}
$\alpha$ is called an extended class $\mathcal{K}_\infty$ function.
\end{defn}

With the aforementioned concepts, the CBF is defined:

\begin{defn}[Control Barrier Function~\cite{DBLP:conf/eucc/AmesCENST19}]
Let $\mathcal{C}\subset\mathcal{D}\subset\mathbb{R}^n$ be the 0-superlevel set of a continuously differentiable function $h: \mathcal{D}\rightarrow\mathbb{R}$, then $h$ is a control barrier function (CBF) on $\mathcal{C}$ if there exists an extended class $\mathcal{K}_\infty$ function $\alpha(\cdot)$ such that for all $\x\in\mathcal{D}$, the system defined in~\eqref{eq:control_affine_system} satisfies
\begin{equation}
    \sup_{\bu\in\mathcal{U}}\Big[\frac{\partial h(\x)}{\partial\x}\Big(\mathbf{f}(\x) + \mathbf{g}(\x)\bu\Big)\Big] \geq -\alpha(h(\x)),
    \label{eq:CBF_constraint}
\end{equation}
with $\alpha: \mathbb{R}\rightarrow\mathbb{R}$ being an extended class $\mathcal{K}_\infty$ function.
\end{defn}

Using the condition in~\eqref{eq:CBF_constraint} and a possibly unsafe performance controller $\pi_\mathrm{perf}: \mathbb{R}^n\rightarrow\mathbb{R}^m$, we can construct a reactive controller by solving a quadratic program (QP) at each time step
\begin{align}
\label{eq:CBFQP}
    \pi(\x) = \argmin_{\bu\in\mathcal{U}}\ &\ \|\bu - \pi_\mathrm{perf}(\x)\|^2\\
    \mathrm{subject\ to}\ &\ \Big[\frac{\partial h(\x)}{\partial\x}\Big(\mathbf{f}(\x) + \mathbf{g}(\x)\bu\Big)\Big] \geq -\alpha(h(\x))\nonumber
\end{align}
which is usually called a CBF-QP~\cite{DBLP:conf/iccps/GurrietSRCFA18}. The CBF-QP can be seen as a safety filter applied on top of $\pi_\mathrm{perf}(\x)$, which finds the closest control in the least-square sense that also enforces forward invariance with respect to $\mathcal{C}$.

\section{Problem Formulation}
\label{sec:formulation}

In this section, we present our assumptions on HCBFs and model uncertainty and define the problem for learning a better CBF under uncertain dynamics. We consider a set of state constraints in the form of 
\begin{equation}
    \mathbf{c}_i(\x) \leq 0,\ i = 1, \cdots, r
    \label{eq:state_constraint}
\end{equation}
where $\mathbf{c}_i: \mathbb{R}^n\rightarrow\mathbb{R}$. We define $\mathcal{S}_i$ as the 0-superlevel set of $-\mathbf{c}_i(\x)$, i.e.,
\begin{equation}
    \mathcal{S}_i = \{\x \mid -\mathbf{c}_i(\x) \geq 0\}.
\end{equation}
We define the intersections of all $\mathcal{S}_i$'s as $\mathcal{S}$, i.e.,
\begin{equation}
    \mathcal{S} = \bigcap_{i=1}^{r}{\mathcal{S}_i}.
\end{equation}
The true safe set $\mathcal{C}$ under the constraints in~\eqref{eq:state_constraint} is defined as the largest forward invariant set contained in $\mathcal{S}$ that can be expressed as the 0-superlevel set of a continuously differentiable function. The notion of forward invariance can be understood as the property that if the control input satisfies~\eqref{eq:CBF_constraint}, then if the initial state of the system $\x_0$ is within the set $\mathcal{C}$, then the state trajectory lies within $\mathcal{C}$ for all $t \in \mathbf{I}(\x_0)$. Thus, we have the relationship
\begin{equation}
    \mathcal{C} \subseteq \mathcal{S}.
\end{equation}
We assume that an unknown continuously differentiable function $h$ is a valid CBF on $\mathcal{C}$. In many cases, even though we cannot directly find a continuously differentiable function with its 0-superlevel set being $\mathcal{C}$, we are able to find another continuously differentiable function $\widehat{h}: \mathbb{R}^n \rightarrow \mathbb{R}$ such that its 0-superlevel set $\widehat{\mathcal{C}}$ is contained within $\mathcal{C}$
\begin{equation}
    \widehat{\mathcal{C}} = \{\x \mid \widehat{h}(\x) \geq 0\} \subseteq \mathcal{C}.
\end{equation}
Assuming that we have access to $\widehat{h}$, without loss of generality, we can write the relationship between $\widehat{h}$ and $h$ as
\begin{equation}
\label{eq:cbf_form}
    h(\x) = \widehat{h}(\x) + \Delta h(\x),
\end{equation}
with $\Delta h: \mathbb{R}^n \rightarrow \mathbb{R}$ being a continuously differentiable function. One assumption we make for $\widehat{h}$ is that it has the same relative degree as $h$\footnote{The system has relative degree $r$ if, in the neighborhood of the equilibrium, $L_\mathbf{g}L_\mathbf{f}^{i-1}\mathbf{j}(\x) = 0$ for $i = 1, 2, \cdots, r-1$ and $L_\mathbf{g}L_\mathbf{f}^{r-1}\mathbf{j}(\x) \neq 0$, where $\mathbf{j}(\x)$ is the output of the system.}. This is a mild assumption~\cite{DBLP:conf/icra/WestenbroekFMAP20}, given that the relative degree of a system represents the actuation capabilities of the system dynamics and can often be inferred from first principles. In this paper, we consider CBFs with relative degree one because, without the loss of generality, we can always use the idea of exponential CBFs~\cite{NguyenS16} to create a CBF with relative degree one starting from a CBF with a higher relative degree.

In the CBF-QP framework, the CBF is not the only source of uncertainty. In practice, the system dynamics in~\eqref{eq:control_affine_system} would be inaccurate because of unmodelled dynamics and parametric errors. Instead of $\mathbf{f}$ and $\mathbf{g}$, we would usually only have access to a nominal model
\begin{equation}
    \dot{\x} = \widehat{\mathbf{f}}(\x) + \widehat{\mathbf{g}}(\x)\bu,
\end{equation}
with locally Lipschitz continuous functions $\widehat{\mathbf{f}}: \mathbb{R}^n\rightarrow\mathbb{R}^n$ and $\widehat{\mathbf{g}}: \mathbb{R}^n\rightarrow\mathbb{R}^{n\times m}$. Similar to the case in CBFs, without loss of generality, we have the relationships
\begin{subequations}
\label{eq:system_dynamics_form}
\begin{align}
    \mathbf{f}(\x) &= \widehat{\mathbf{f}}(\x) + \Delta\mathbf{f}(\x),\\
    \mathbf{g}(\x) &= \widehat{\mathbf{g}}(\x) + \Delta\mathbf{g}(\x),
\end{align}
\end{subequations}
with locally Lipschitz continuous functions $\Delta\mathbf{f}: \mathbb{R}^n\rightarrow\mathbb{R}^n$ and $\Delta\mathbf{g}: \mathbb{R}^n\rightarrow\mathbb{R}^{n\times m}$. We assume that the nominal dynamics have the same relative degree as the true dynamics, which is a common assumption in the literature~\cite{TaylorSYA20}\cite{WangMLS021}. Since we only have a conservative estimation of $h$ and the nominal dynamics, if we deploy CBF-QP using these known functions, there would be no safety guarantees. Thus, the main goal of this paper is to find an algorithmic approach to learning the functions $\Delta h$, $\Delta\mathbf{f}$, and $\Delta\mathbf{g}$, which will be discussed in Section~\ref{sec:method}.
\section{Method}
\label{sec:method}

In this section, we propose an algorithmic approach to solve the problem formulated in Section~\ref{sec:formulation}. The structure of this section is as follows. First, we describe our proposed solution to the CBF learning problem. Then, we describe how we learn the system dynamics. Finally, we show how we jointly solve these two learning problems.

\subsection{Learning the Control Barrier Function}
\begin{figure}[t!]
    \centering
    \includegraphics[width=0.49\textwidth]{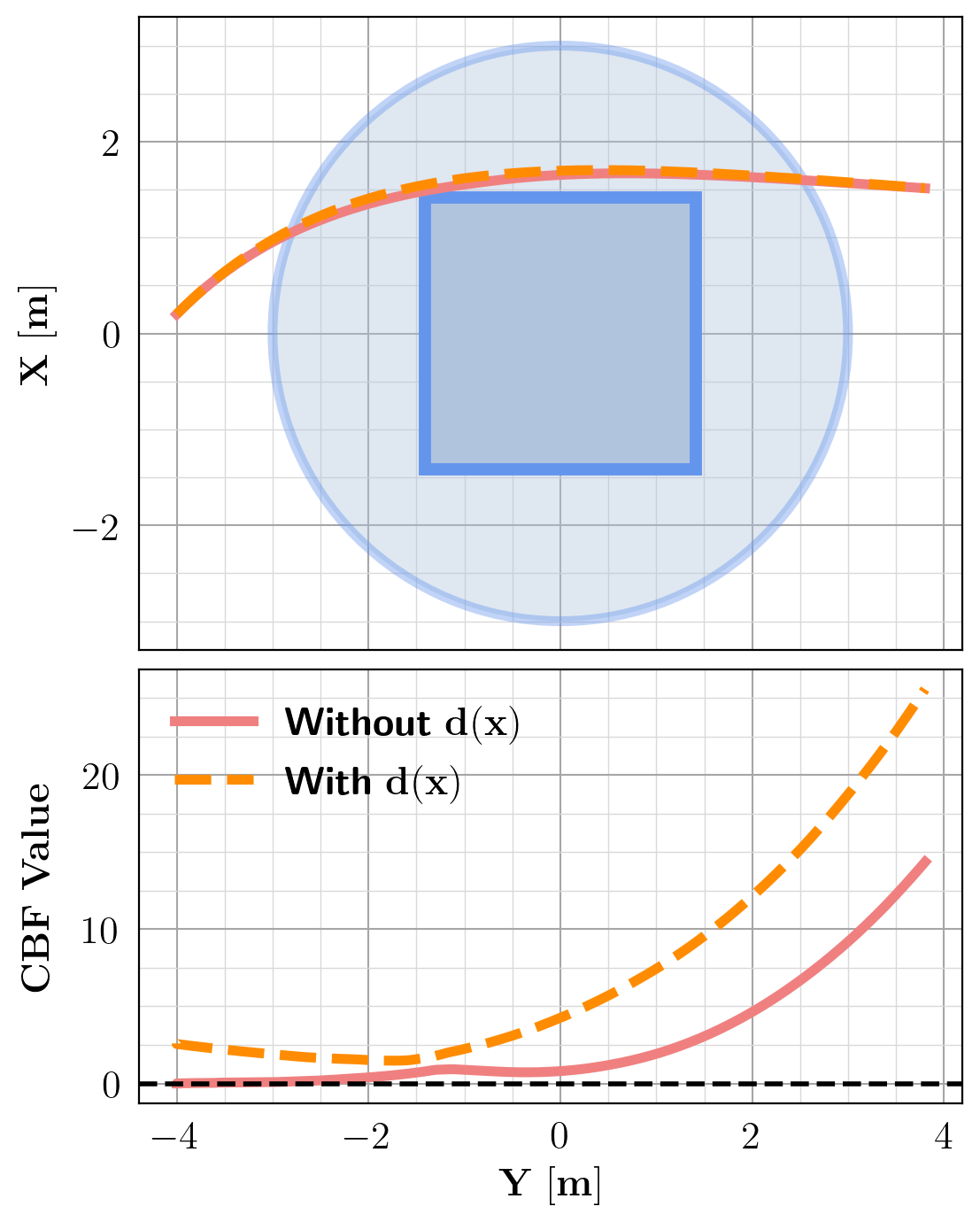}
    \caption{Illustration of the effect of the distance function $\mathbf{d}(\mathbf{x})$. These figures show the result of learning a CBF for a target-reaching-obstacle-avoidance task on a unicycle (see Section~\ref{sec:unicycle} for details). In the upper graph, the light blue region represents the unsafe set under the HCBF, and the darker blue region represents the obstacle. The upper graph shows the trajectory generated by a CBF-QP controller with a proportional controller (see Section~\ref{sec:unicycle} for details) as the performance controller and using the learned CBF-QP. The lower graph shows the evolution of the CBF value along the trajectory.}
    \label{fig:unicycle_dx}
\end{figure}

\begin{figure*}[t!]
    \centering
    \includegraphics[width=\textwidth]{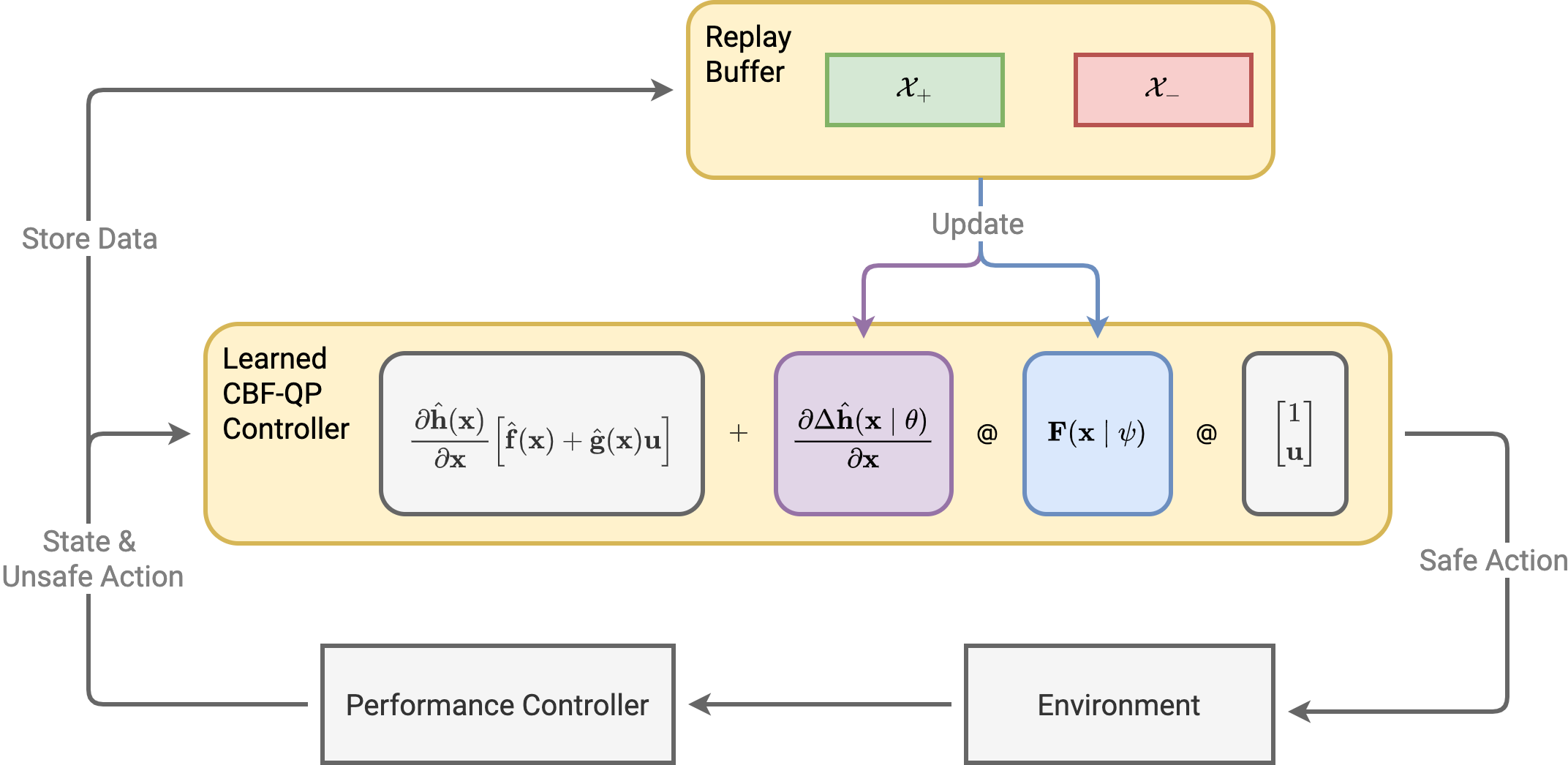}
    \caption{Illustration of the overall training procedure. The ``@" sign represents matrix multiplication. The gray boxes represent the non-learning components. The purple box represents the learned CBF, the blue box represents the learned system dynamics, the green box represents the buffer storing safe interactions, and the red box represents the buffer storing unsafe interactions.}
    \label{fig:LearnCBFUncertain}
\end{figure*}

Following the problem formulation in Section~\ref{sec:formulation}, we need to estimate $\Delta h(\x)$ in order to estimate the CBF. We propose to use a deep neural network (DNN) to estimate $\Delta h(\x)$, we write this DNN as $\Delta\widehat{h}(\x\mid\theta)$, where $\theta$ represents the weights of the DNN. 

Given that $\Delta h(\x)$ is a continuously differentiable function, we also require $\Delta\widehat{h}(\x\mid\theta)$ to be a continuously differentiable function with respect to $\x$. To achieve this, we use a deep differential network with smooth activation functions, which we refer the reader to~\cite{DBLP:conf/iclr/LutterRP19} for a detailed description. The deep differential network has two forward paths. One of the paths is the same as in standard fully-connected DNNs. The other computes the Jacobian of the DNN with respect to its input. Since it directly outputs the Jacobian, compared to performing an additional numerical differentiation pass, using deep differential networks increases the computational efficiency. A single layer within a deep differential network has the form
\begin{equation}
    (\frac{\partial\bar{\mathbf{y}}}{\partial\mathbf{y}},\ \bar{\mathbf{y}}) = \ell(\mathbf{y}),
\end{equation}
where $\mathbf{y}\in\mathbb{R}^{n_i\times1}$ is the input of the layer, $\bar{\mathbf{y}}\in\mathbb{R}^{n_o\times1}$ is the output of the layer, and the layer is represented by $\ell: \mathbb{R}^{n_i\times1} \rightarrow (\mathbb{R}^{n_o \times n_i}, \mathbb{R}^{n_o\times1})$. The Jacobian is computed as
\begin{equation}
    \frac{\partial\bar{\mathbf{y}}}{\partial\y} = \mathrm{diag}(\mathbf{g}^\prime(\mathbf{a}))\mathbf{W},
\end{equation}
where $\mathbf{W}\in\mathbb{R}^{n_o\times n_i}$ represents the weights of that layer, $\mathbf{g}: \mathbb{R}^{n_o\times1}\rightarrow\mathbb{R}^{n_o\times1}$ represents the activation function, $\mathbf{g}^\prime(\cdot)$ represents the derivative of the activation function, and $\mathbf{a} = \mathbf{W}\y + \mathbf{bias}$.

To find the weights $\theta$, we would need to collect a dataset of features and labels. However, since we do not have access to a CBF with its 0-superlevel set coinciding with $\mathcal{C}$, we do not have groundtruth labels. A widely used approach~\cite{DBLP:conf/cdc/RobeyHLZDTM20}~\cite{DBLP:journals/ral/QinSF22} is to learn a valid CBF without groundtruth labels by utilizing the properties in~\eqref{eq:CBF_conditions} and write the loss functions for learning $\theta$ as
\begin{subequations}
\label{eq:old_loss_functions}
\begin{align}
    \mathcal{L}_+(\theta) &= \frac{1}{N}\sum_{\x_i\in\mathcal{X}_+}^{}\max\Big(0, -\tilde{h}(\x_i\mid\theta)\Big),\\
    \mathcal{L}_-(\theta) &= \frac{1}{N}\sum_{\x_i\in\mathcal{X}_-}^{}\max\Big(0, \tilde{h}(\x_i\mid\theta)\Big),
\end{align}
\end{subequations}
with $\mathcal{L}_+$ representing the loss for safe states, $\mathcal{L}_-$ representing the loss for unsafe states, $\mathcal{X}_{+}$ being the dataset containing safe interactions (state-control pair), $\mathcal{X}_{-}$ being the dataset containing unsafe interactions, and
\begin{equation}
\label{eq:estimated_CBF}
    \tilde{h}(\x\mid\theta) = \widehat{h}(\x) + \Delta\widehat{h}(\x\mid\theta).
\end{equation}
We can see that for a safe state, $\mathcal{L}_+$ is only non-zero when the estimated CBF $\widehat{h}(\x) + \Delta\widehat{h}(\x\mid\theta)$ is negative. For unsafe states, $\mathcal{L}_-$ is only non-zero when the estimated CBF is positive. Thus, in both cases, only when the sign of the estimated CBF is wrong will there be a non-zero loss; otherwise, the loss is zero. However, one trivial solution for $\Delta\widehat{h}$ that minimizes the losses is
\begin{equation}
    \Delta\widehat{h}(\x) = -\widehat{h}(\x).
\end{equation}
While this is a minimizer for both $\mathcal{L}_+$ and $\mathcal{L}_-$, it will also make the estimated CBF zero everywhere, making it an undesired solution. Although, when combined with the loss derived from the CBF constraint (which will be discussed later in this section), the learned CBF will not constantly be zero, it will be close to zero for a large portion of the state space, which makes it difficult to distinguish between safe and unsafe states. This phenomenon can be seen in Fig.~\ref{fig:unicycle_dx}, where the ``Without $\mathbf{d}(x)$" case is trained using the losses in~\eqref{eq:old_loss_functions}.  

To deal with this issue, we note that the sign and trend of the CBF matters while its magnitude is of less importance. Using this intuition, we establish a simple heuristic, i.e., the further outside the safe set, the more negative the CBF value should be, and the more inside the safe set, the more positive the CBF value should be. We define the notion of ``more inside" and ``more outside" using the state constraints in~\eqref{eq:state_constraint}. For a single constraint $\mathbf{c}(\x) \leq 0$, we can compute the value of
\begin{equation}
    \mathbf{d}(\x) = -\mathbf{c}(\x).
    \label{eq:distance_function}
\end{equation}
When $\mathbf{d}(\x)$ is positive, the larger it is, the more inside the safe set $\x$ is. When $\mathbf{d}(\x)$ is negative, the smaller it is, the more outside the safe set $\x$ is. When there are multiple constraints, we can compute
\begin{equation}
    \mathbf{d}_i(\x) = -\mathbf{c}_i(\x),\ \forall i = 1,\ \cdots,\ r.
\end{equation}
Then, $\mathbf{d}(\x)$ is defined as
\begin{equation}
    \mathbf{d}(\x) = \min\{\mathbf{d}_1(\x),\ \cdots, \mathbf{d}_r(\x)\}.
\end{equation}
Using the $\mathbf{d}(\x)$'s, we can write the new loss functions as
\begin{subequations}
\begin{align}
    \mathcal{L}_+(\theta) &= \frac{1}{N}\sum_{\x_i\in\mathcal{X}_+}^{}\max\Big(0, -\tilde{h}(\x_i\mid\theta) + \mathbf{d}_{+}(\x_i)\Big)\\
    \mathcal{L}_-(\theta) &= \frac{1}{N}\sum_{\x_i\in\mathcal{X}_-}^{}\max\Big(0, \tilde{h}(\x_i\mid\theta) - \mathbf{d}_{-}(\x_i)\Big),
\end{align}
\end{subequations}
where $\mathbf{d}_{+}, \mathbf{d}_{-}: \mathbb{R}^n \rightarrow \mathbb{R}$ represents the distance functions corresponding to the safe and unsafe set, respectively. The effect of having $\mathbf{d}(\x)$ in the loss function can be seen in Fig.~\ref{fig:unicycle_dx} (``With $\mathbf{d}(\x)$" curve), where the CBF value is no longer flat near the obstacle. 

In addition to the CBF losses, we add another loss corresponding to the CBF constraint in~\eqref{eq:CBF_constraint} to ensure the ability to generate safe control actions
\begin{align}
    \mathcal{L}_\mathbf{\nabla h}(\theta) =&\ \frac{1}{N}\sum_{\x_i\in\mathcal{X}_+}^{}\max\Big(0, \frac{\partial\tilde{h}(\x_i\mid\theta)}{\partial\x}\dot{\tilde{\x}}_i\nonumber\\
    &- \alpha(\tilde{h}(\x_i\mid\theta))\Big),
\end{align}
where $\dot{\tilde{\x}}_i$ is modeled using the learned system dynamics (see Section~\ref{sec:learn_system_dynamics}). Although the learned system dynamics would also be parameterized by a set of weights, when performing gradient-based updates, the gradient of $\mathcal{L}_\mathbf{\nabla h}$ is only calculated with respect to $\theta$. We will defer the discussion of the learning procedure to Section~\ref{sec:training_process}. We also add a term in our loss function to regulate the amount of change in $\tilde{h}(\x\mid\theta)$ induced by $\Delta\widehat{h}(\x\mid\theta)$ as
\begin{equation}
    \mathcal{L}_{\Delta h}(\theta) = \frac{1}{N}\sum_{\x_i\in\mathcal{X}_+\cup\mathcal{X}_-}^{}{\Delta\widehat{h}^2(\mathbf{x}_i\mid\theta)}.
\end{equation}
By weighting this term against the other terms in the loss function, we can add a prior on how confident the user is in the ability of the HCBF to recover the safe set.

Using the terms defined above, the final loss function is given by
\begin{equation}
\label{eq:theta_loss}
    \mathcal{L}_\theta(\theta) = \mathcal{L}_+ + \lambda_1\mathcal{L}_- + \mathcal{L}_\mathbf{\nabla h} + \lambda_2\mathcal{L}_{\Delta h},
\end{equation}
with $\lambda_1, \lambda_2\in\mathbb{R}_+$ weighting the importance of the individual loss terms. Since estimating part of the unsafe set as safe is much more disastrous than estimating part of the safe set as unsafe, $\lambda_1$ is usually larger than one.

\subsection{Learning the System Dynamics}
\label{sec:learn_system_dynamics}
To learn the system dynamics, we use another neural network parameterized by $\psi$, i.e., $\mathbf{F}(\x\mid\psi)$, to estimate both $\Delta\mathbf{f}(\x)$ and $\Delta\mathbf{g}(\x)$. Using this neural network, our estimated dynamics is defined as
\begin{equation}
    \dot{\tilde{\x}}(\x, \bu\mid\psi) = \widehat{\mathbf{f}}(\x) + \widehat{\mathbf{g}}(\x)\bu + \mathbf{F}(\x\mid\psi)\begin{bmatrix}
        1\\
        \bu
    \end{bmatrix}.
    \label{eq:estimated_dynamics}
\end{equation}
Common methods in learning the system dynamics require obtaining data of $\dot{\x}$~\cite{DBLP:conf/iclr/LutterRP19} or the next state (i.e., state at the ``next'' time step)~\cite{DBLP:conf/icra/WestenbroekFMAP20}. Using $\dot{\x}$ requires additional sensors, e.g., inertial measurement units. Using the next state is also not accurate, since the commonly used integration schemes only approximate the true discrete-time dynamics. Thus, instead of learning the system dynamics via a regression problem on $\dot{\x}$ or the next state, we form a regression problem on $\dot{h}(\x)$~\cite{TaylorSYA20}, which is given by
\begin{equation}
    \dot{h}(\x) = \frac{\partial h(\x)}{\partial\x}\Big(\mathbf{f}(\x) + \mathbf{g}(\x)\bu\Big).
\end{equation}
Using the estimated dynamics from~\eqref{eq:estimated_dynamics} and the estimated CBF from~\eqref{eq:estimated_CBF}, the estimated $\dot{h}(\x)$ is given as
\begin{equation}
    \dot{\tilde{h}}(\x, \bu, \theta \mid \psi) = \frac{\partial\tilde{h}(\x, \theta)}{\partial\x}\dot{\tilde{\x}}(\x, \bu\mid\psi).
\end{equation}
When learning the system dynamics, although the value of $\dot{\tilde{h}}$ does depend on $\theta$, the gradient is only calculated with respect to $\psi$. Therefore, in the remainder of this section, we will omit $\dot{\tilde{h}}$'s dependency on $\bu$ and $\theta$. Additionally, we can numerically estimate $\dot{h}(\x)$ using the central difference method
\begin{equation}
    \widehat{\dot{\tilde{h}}}(\x) = \frac{\tilde{h}(\x_+) - \tilde{h}(\x_-)}{2\Delta{t}},
\end{equation}
where $\x_+$ represents the next state, and $\x_-$ represents the previous state. To learn the weights $\psi$, we use the loss function defined as
\begin{equation}
\label{eq:loss_system_dynamics}
    \mathcal{L}_\psi(\psi) = \sum_{\x_i\in\mathcal{X}_+\cup\mathcal{X}_-}\Big(\widehat{\dot{\tilde{h}}}(\x_i) - \dot{\tilde{h}}(\x_i\mid\psi)\Big)^2.
\end{equation}
Note that when performing gradient-based updates for $\psi$, the gradient of $\mathcal{L}_\psi$ is only calculated with respect to $\psi$. Given that
\begin{equation}
    \lim_{\Delta{t}\rightarrow0}\widehat{\dot{\tilde{h}}}(\x) = \frac{\partial\tilde{h}(\x)}{\partial\x}\dot{\x},
\end{equation}
we can show that for a small $\Delta{t}$ and loss value, the error in the learned dynamics is bounded. Assuming the loss is less than some positive value, i.e., 
\begin{equation}
    \mathcal{L}_\psi(\psi) \leq \epsilon,
\end{equation}
where $\epsilon\in\mathbb{R}_+$, yields
\begin{equation}
    \widehat{\dot{\tilde{h}}}(\x_i) - \dot{\tilde{h}}(\x_i\mid\psi) \leq \sqrt{\epsilon}.
\end{equation}
Given that the central difference method has a truncation error of $\mathcal{O}(\Delta{t}^2)$, we have
\begin{equation}
    \widehat{\dot{\tilde{h}}}(\x) = \frac{\partial\tilde{h}(\x)}{\partial\x}\dot{\x} + \mathcal{O}(\Delta{t}^2),
\end{equation}
which leads to
\begin{equation}
    \frac{\partial\tilde{h}(\x_i)}{\partial\x}\Big(\dot{\x}_i - \dot{\tilde{\x}}(\x_i, \bu\mid\psi^*)\Big) + \mathcal{O}(\Delta{t}^2) \leq \sqrt{\epsilon}.
\end{equation}
Then, we have the following bound on the error of the learned dynamics, i.e., $\dot{\x}_i-\dot{\tilde{\x}}$:
\begin{equation}
    \dot{\x}_i - \dot{\tilde{\x}}(\x_i, \bu\mid\psi^*) \leq \displaystyle\Big[\frac{\partial\tilde{h}(\x_i)}{\partial\x}\Big]^{\dagger}\Big(\sqrt{\epsilon} + \mathcal{O}(\Delta{t}^2)\Big).
\end{equation}
This shows that with a small enough $\Delta{t}$ and loss value, our  proposed algorithm can learn a reasonably accurate model of the system dynamics.

\subsection{Training Process}
\label{sec:training_process}
We train $\Delta\widehat{h}$ and $\mathbf{F}$ using a supervised learning approach. For supervised learning, one key assumption for the training data is that they are independently and identically distributed (i.i.d). Thus, instead of only training the networks using data collected from the current episode, we store the data in replay buffers~\cite{DBLP:journals/nature/MnihKSRVBGRFOPB15} and only use randomly sampled data from the replay buffer to train the network. We form two replay buffers, one for safe data $\mathcal{X}_+$ and one for unsafe data $\mathcal{X}_-$.

The overall training procedure is as follows. At each time step, given the current state, the performance controller computes a potentially unsafe action $\bu_{\mathrm{perf}}(\x)$. Then, the unsafe action is passed through the learned CBF filter, making it the estimated safe action. The learned CBF-QP controller has the form
\begin{align}
    \min_{\bu\in\mathcal{U}}\ &\ \|\bu - \bu_{\mathrm{perf}}(\x)\|\\
    \mathrm{subject\ to}\ &\ \frac{\partial\tilde{h}(\x)}{\partial\x}\dot{\tilde{\x}}(\x, \bu\mid\psi) \geq -\alpha(\tilde{h}(\x))\nonumber.
\end{align}
Finally, the control action $\bu$ is applied to the environment. Additionally, the current state, the learned CBF, and the estimated safe action are stored in the corresponding replay buffer at each time step. After each episode ends, data sampled from the replay buffer are used to compute the loss functions in~\eqref{eq:theta_loss} and~\eqref{eq:loss_system_dynamics}. Then, using a stochastic gradient descent algorithm, e.g., ADAM~\cite{DBLP:journals/corr/KingmaB14}, the weights of the two networks $\Delta\widehat{h}$ and $\mathbf{F}$ are updated. This procedure is repeated until the two networks converge or if a predefined maximum episode number is reached. A visual illustration of this procedure can be found in Fig.~\ref{fig:LearnCBFUncertain}. For our proposed approach, all of the learning is done offline, either in a simulation environment or a specially designed experiment environment. After the learning process converges, the learned CBF-QP can then be deployed to the intended system.
\section{Simulation Studies}
In this section, we show the effectiveness of our approach using three systems: double integrator, unicycle, and two-link arm. All experiments are performed using PyTorch with the same neural network architecture. The deep differential network $\widehat{h}$ consists of three layers with output sizes $[128, 128, 1]$. The dynamics network $\mathbf{F}$ consists of two networks, one for estimating $\Delta{\mathbf{f}}$, with output size $[64, 64, n]$, and the other estimates $\Delta{\mathbf{g}}$, with output size $[64, 64, nm]$, which is reshaped as a $n\times m$ matrix.

\subsection{Double Integrator}
\begin{figure}[t!]
    \centering
    \includegraphics[width=0.49\textwidth]{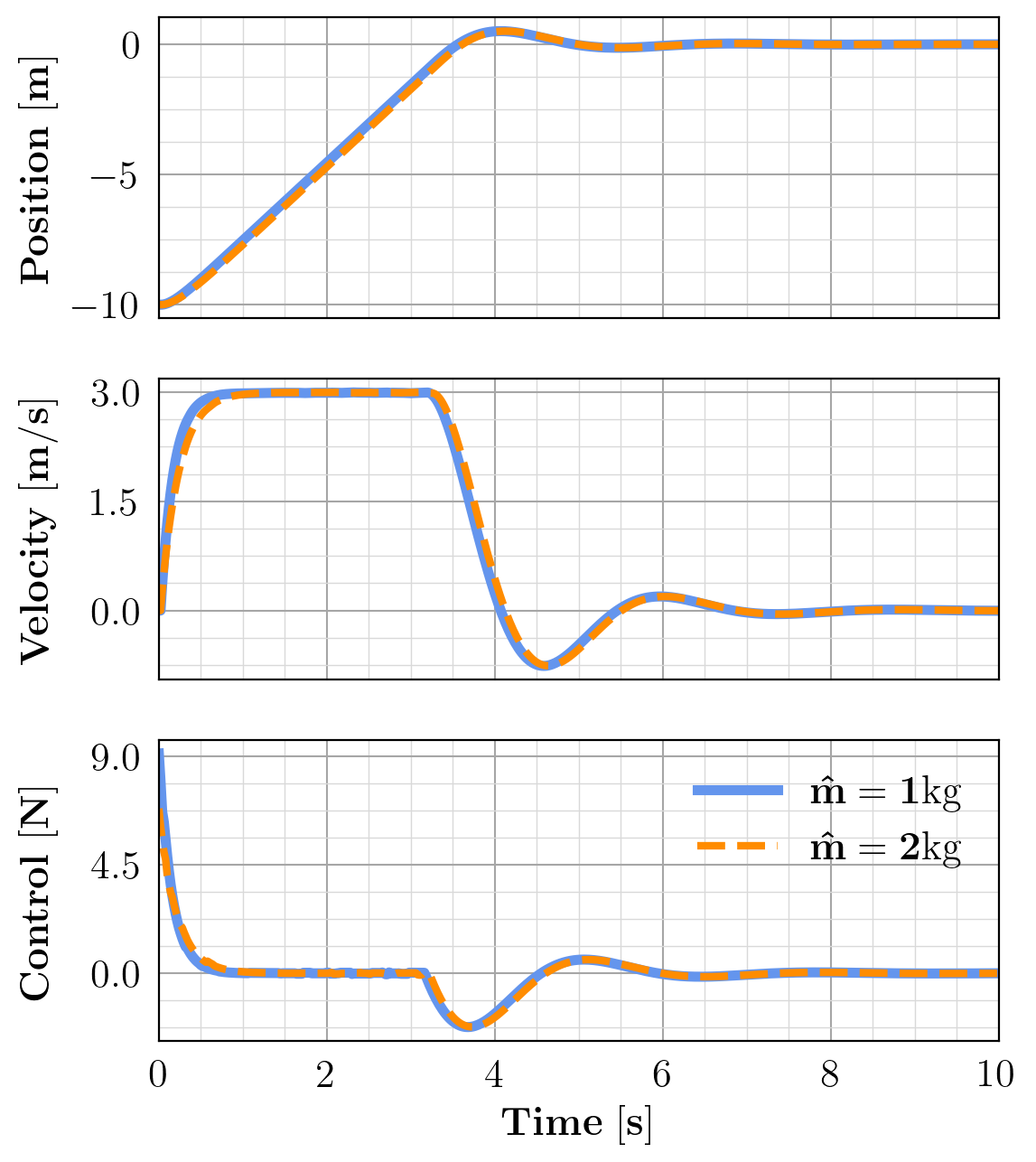}
    \caption{Illustration of the state and control trajectory for the double integrator system. The learned CBF-QP controller generates the trajectories under two different initial guesses of the system dynamics.}
    \label{fig:integrator2dmass}
\end{figure}

\begin{figure}[t!]
    \centering
    \includegraphics[width=0.49\textwidth]{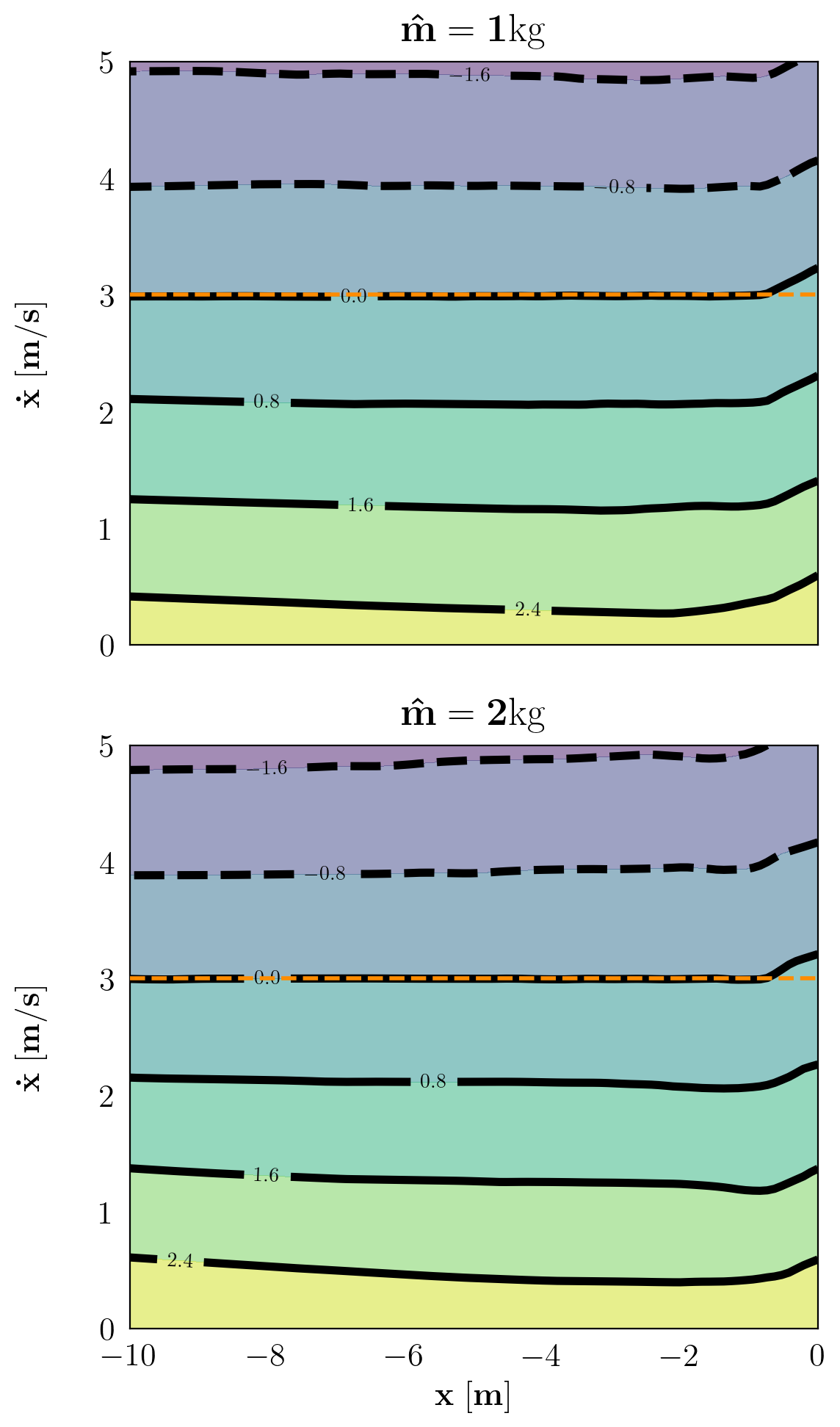}
    \caption{Contour of the learned CBF for the double integrator system. The orange dashed line denotes the zero-level line of the true CBF. The region above the orange dashed line should be negative for the true CBF and positive below. The contour values correspond to the CBF level sets.}
    \label{fig:integrator2dmass_cbf_contour}
\end{figure}

\begin{figure}
    \centering
    \includegraphics[width=0.49\textwidth]{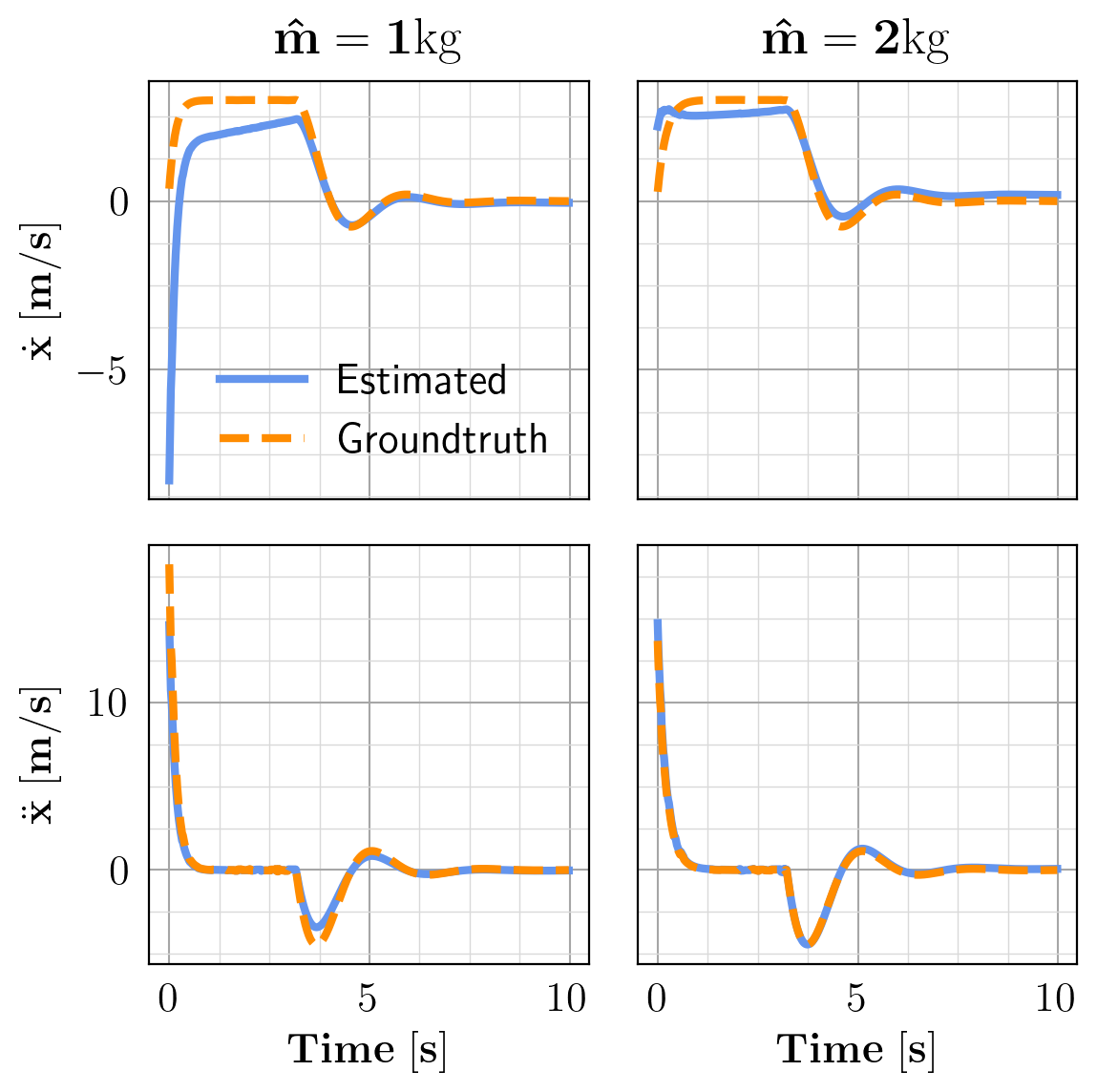}
    \caption{Comparison between the trajectories generated by the estimated (learned) and groundtruth dynamics starting from different initial guesses for the double integrator system.}
    \label{fig:integrator2dmass_model_est}
\end{figure}

The double integrator has the system dynamics given as
\begin{equation}
\label{eq:2dintegrator_dynamics}
    \begin{bmatrix}
        \dot{x}\\
        \ddot{x}
    \end{bmatrix} = \begin{bmatrix}
        0 & 1\\
        0 & 0
    \end{bmatrix}\begin{bmatrix}
        x\\
        \dot{x}
    \end{bmatrix} + \begin{bmatrix}
        0\\
        1/m
    \end{bmatrix}\bu,
\end{equation}
with $x\in\mathbf{R}$ denoting the position, $\dot{x}\in\mathbf{R}$ denoting the velocity, $\bu\in\mathbb{R}$ denoting the control, and $m\in\mathbf{R}_+$ denoting the mass. In our simulation environment, we set $m = 0.5$kg, however, we assume that $m$ is unknown. The system has a velocity constraint
\begin{equation}
    \dot{x} \leq 3.
\end{equation}
We construct the HCBF as
\begin{equation}
    \widehat{h}(\x) = 2 - \dot{x},
\end{equation}
which corresponds to the constraint
\begin{equation}
    \dot{x} \leq 2.
\end{equation}
Since this is a simple example, we can also get one of the CBFs that recovers the entire safe set
\begin{equation}
    h(\x) = 3 - \dot{x},
\end{equation}
which can be used to check the quality of the learned CBF. During training, we use a PD controller as the performance controller
\begin{equation}
    \pi(\x) = \mathbf{K}_p(x_{\mathrm{des}} - x) + \mathbf{K}_d(\dot{x}_{\mathrm{des}} - \dot{x}),
\end{equation}
with $\mathbf{K}_p = 3$, $\mathbf{K}_d = 1.0$, and $[x_{\mathrm{des}}, \dot{x}_{\mathrm{des}}] = [0, 0]$. During training, we set $\lambda_1 = 100.0$ and $\lambda_2 = 1.0$. The learning rate is $10^{-4}$. We set
\begin{equation}
    \mathbf{d}_+(\x) = \mathbf{d}_-(\x) = 3 - \dot{x}.
\end{equation}
The class $\mathcal{K}_\infty$ function $\alpha$ is set to be
\begin{equation}
\label{eq:gamma_func}
    \alpha(\x) = \gamma\x.
\end{equation}
During training, the initial state of the system is uniformly sampled with $x_0\in[-15, -5]$ and $\dot{x}_0 = 0$. We provide an initial guess of the system dynamics by replacing the $m$ in~\eqref{eq:2dintegrator_dynamics} with our guess $\widehat{m}$. Using our proposed algorithm, we trained for 100 epochs, and the trajectory generated by the learned CBF-QP controller is shown in Fig.~\ref{fig:integrator2dmass}. It can be seen that even though the initial guess is different, the trajectories generated by the learned CBF-QP controller are very similar. Additionally, the state trajectories are safe, despite the errors in the HCBF and the nominal dynamics. The contour plot of the learned CBF is shown in Fig.~\ref{fig:integrator2dmass_cbf_contour}. It can be seen that the learned safe set almost recovers the true safe set, except for $x$ values near zero. This is due to having little training data where the $x$ values are close to zero and $\dot{x}$ near $3$m/s. A comparison between the learned (estimated) and groundtruth dynamics is shown in Fig.~\ref{fig:integrator2dmass_model_est}. We can see that the learned dynamics are invariant to the initial guess and provides a relatively accurate estimation of the groundtruth dynamics.

\subsection{Unicycle}
\label{sec:unicycle}

\begin{figure*}[t!]
    \centering
    \includegraphics[width=0.92\textwidth]{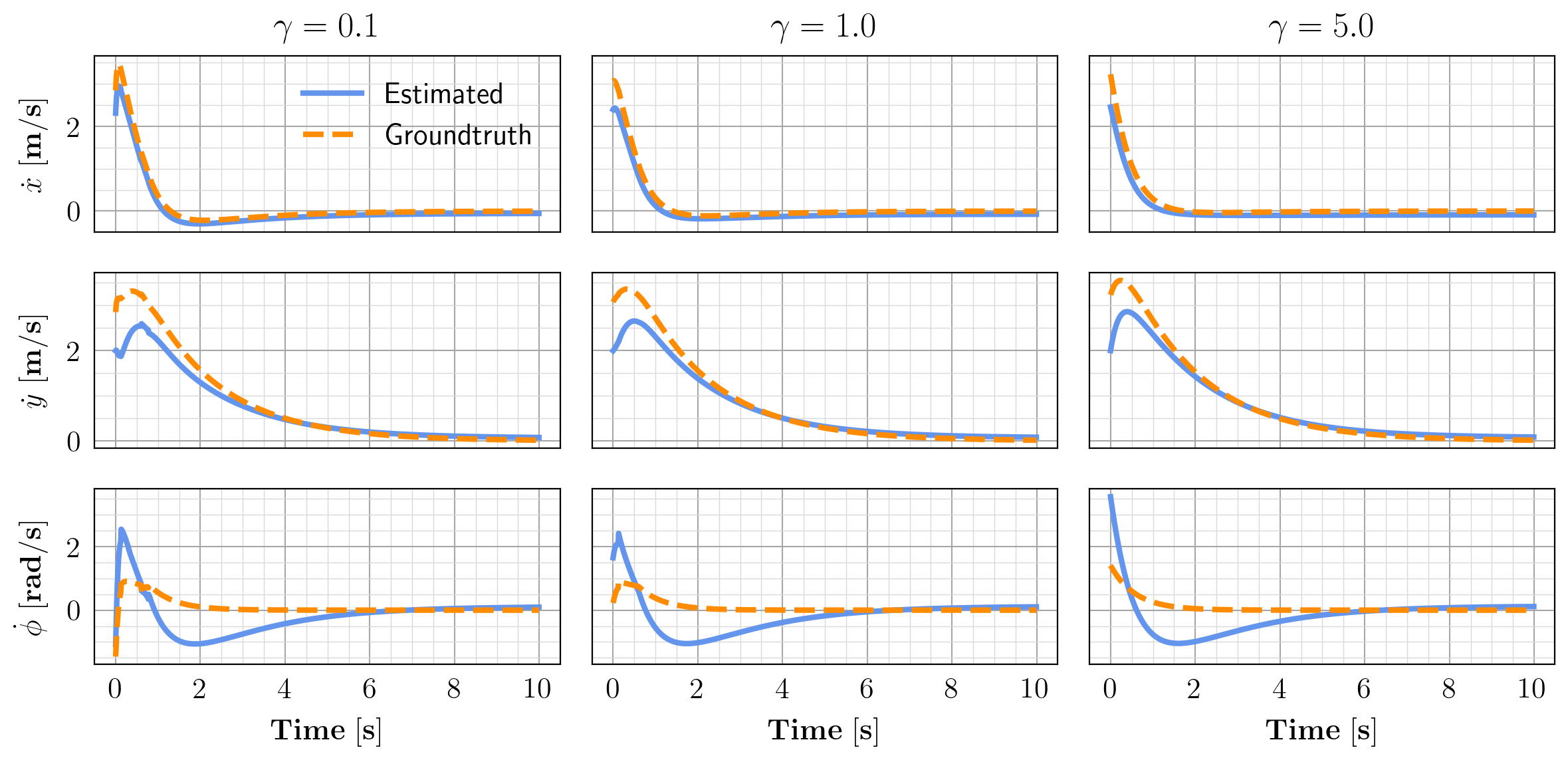}
    \caption{Comparison between the state trajectories of the groundtruth dynamics and the estimated (learned) dynamics for the unicycle system under different values of $\gamma$. For larger $\gamma$ values, the CBF-QP generates control actions that approach the boundary of the safe set more aggressively.}
    \label{fig:unicycle_model_est}
\end{figure*}

\begin{figure}[t!]
    \centering
    \includegraphics[width=0.49\textwidth]{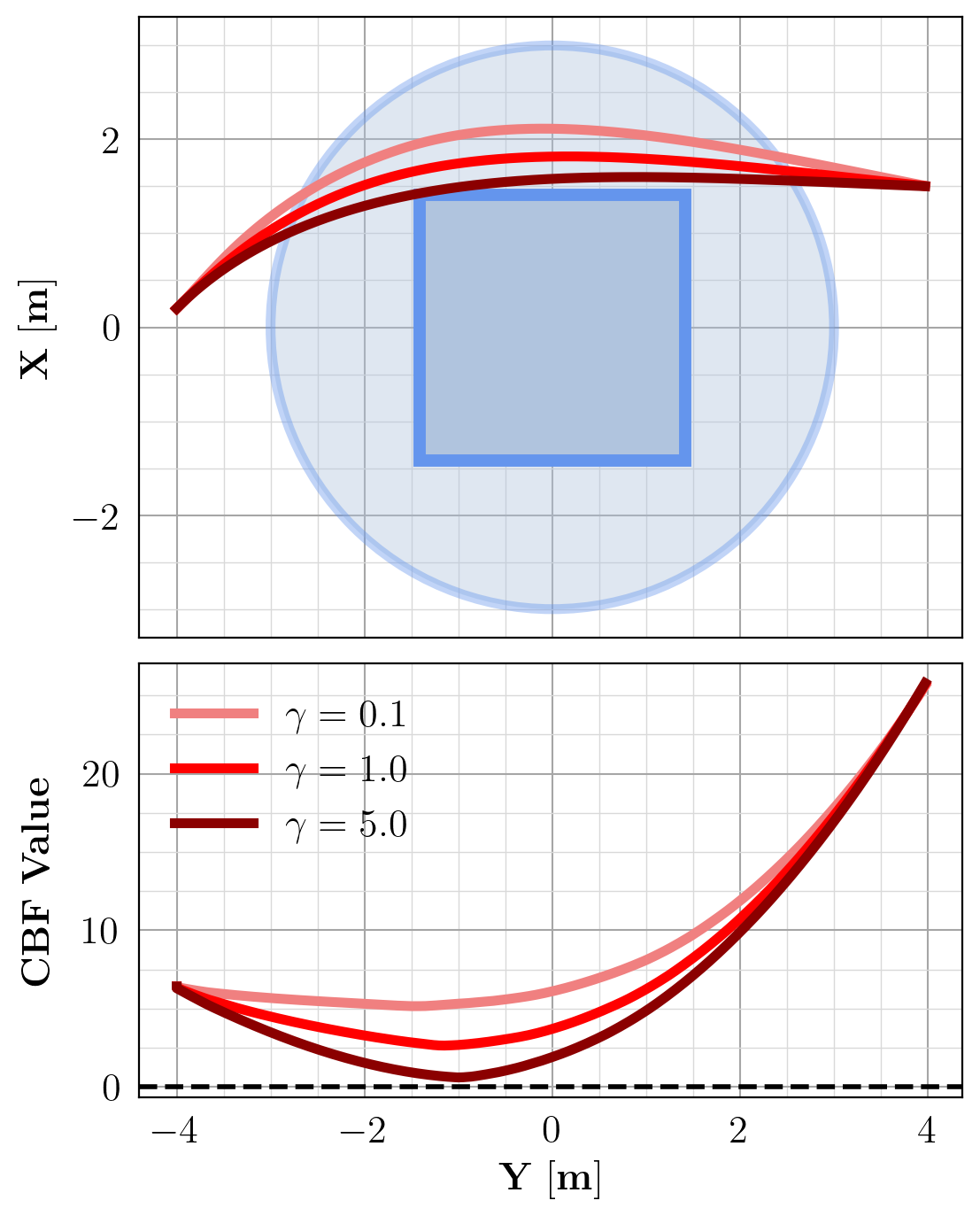}
    \caption{The upper figure shows the motion generated by the learned CBF-QP controller for the unicycle system under different values of $\gamma$. The darker blue square represents the square obstacle. The light blue circle represents the unsafe region corresponding to the HCBF. The lower figure shows the CBF values along the trajectories for different values of $\gamma$.}
    \label{fig:unicycle}
\end{figure}

\begin{figure}[t!]
    \centering
    \includegraphics[width=0.49\textwidth]{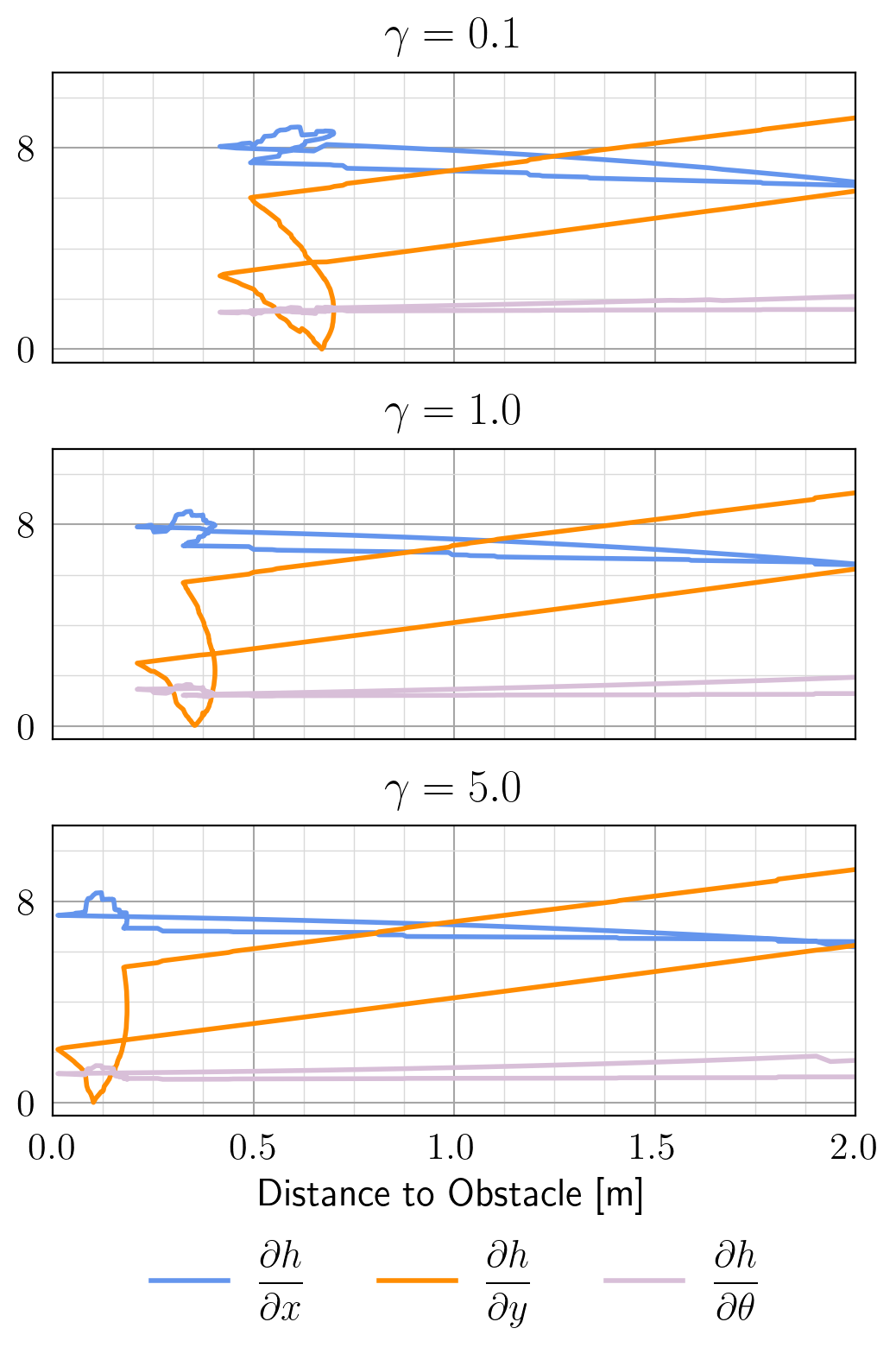}
    \caption{Illustration of how the partial derivative of the learned CBF changes with respect to the distance between the unicycle and the obstacle.}
    \label{fig:dhdx_dynamics_error_reasoning}
\end{figure}

The unicycle system has the dynamics
\begin{equation}
\label{eq:unicycle_dynamics}
    \begin{bmatrix}
        \dot{x}\\
        \dot{y}\\
        \dot{\phi}
    \end{bmatrix} = \begin{bmatrix}
        \alpha_v\cos\phi & 0\\
        \alpha_v\sin\phi & 0\\
        0 & \alpha_\omega
    \end{bmatrix}\begin{bmatrix}
        v\\
        \omega
    \end{bmatrix},
\end{equation}
with $x$ denoting the position of the unicycle along the $x$ axis, $y$ denoting the position along the $y$ axis, and $\phi$ denoting the heading of the unicycle. The terms $\alpha_v$ and $\alpha_\omega$ regulate how the control input $v\in\mathbb{R}$ and $\omega\in\mathbb{R}$ affect the velocity and angular velocity of the unicycle, respectively. We assume that the values of $\alpha_v$ and $\alpha_\omega$ are unknown. The system performs an obstacle avoidance task, where the obstacle is a square. Compared to a square, it is easier to write an HCBF for a circular obstacle~\cite{DBLP:journals/corr/abs-2110-05415} as
\begin{equation}
    \widehat{h}(\x) = x^2 + y^2 + 2xl\cos\phi + 2yl\sin\phi + l^2 - r^2,
\end{equation}
where $r$ is the radius of the constructed circular obstacle and $l$ is a predefined lookahead distance. This choice of HCBF corresponds to the state constraint
\begin{equation}
    x^2 + y^2 \geq r^2.
\end{equation}
This setup is shown in Fig.~\ref{fig:unicycle}. During training, we use a proportional controller as the performance controller
\begin{equation}
    \pi(\x) = \begin{bmatrix}
        \mathbf{K}_ve\\
        \mathbf{K}_\omega(\beta - \phi)
    \end{bmatrix},
\end{equation}
where $\mathbf{K}_v = 0.75$, $\mathbf{K}_\omega = 3.0$, and 
\begin{subequations}
\begin{align}
    e &= \sqrt{(x - x_{\mathrm{des}})^2 + (y - y_{\mathrm{des}})^2}\\
    \beta &= \mathrm{atan}2(y_{\mathrm{des}} - y, x_{\mathrm{des}} - x).
\end{align}
\end{subequations}
For the loss parameters, we set $\lambda_1 = 10.0$ and $\lambda_2 = 0.0$. The learning rate is set to be $10^{-5}$ and
\begin{equation}
    \mathbf{d}_{+}(\x) = \mathbf{d}_{-}(\x) = \max(|x|, |y|) - \ell_s/2.
\end{equation}
where $\ell_s$ represents the side length of the square. The class $\mathcal{K}_\infty$ function $\alpha$ is the same as in~\eqref{eq:gamma_func}. The system is trained for 500 epochs, and the trajectory generated by the learned CBF-QP controller is shown in Fig.~\ref{fig:unicycle}. During training, we set $\gamma = 5.0$. It can be seen that after training if we change the value of $\gamma$, we can still generate safe trajectories. Furthermore, as $\gamma$ gets smaller, the controller gets more conservative, which is the expected behavior. This shows that even using the learned CBF, we can tune the performance of the controller without additional training. The difference between the estimated and learned system dynamics is shown in Fig.~\ref{fig:unicycle_model_est}. The groundtruth values for the control regulation terms are $\alpha_v = 0.75$ and $\alpha_\omega = 0.75$; our initial guess is $\alpha_v = 1.0$ and $\alpha_\omega = 1.0$. As we can see, the learned (estimated) dynamics are different in many cases from the groundtruth dynamics. However, the safety of the learned CBF-QP controller is not violated. As shown in Fig.~\ref{fig:dhdx_dynamics_error_reasoning}, the partial derivative of the learned CBF with respect to $x$ is larger than the other two elements, which makes estimation errors in $\dot{y}$ and $\dot{\phi}$ less significant. 

\subsection{Two-Link Arm}
\begin{figure}[t!]
    \centering
    \includegraphics[width=0.49\textwidth]{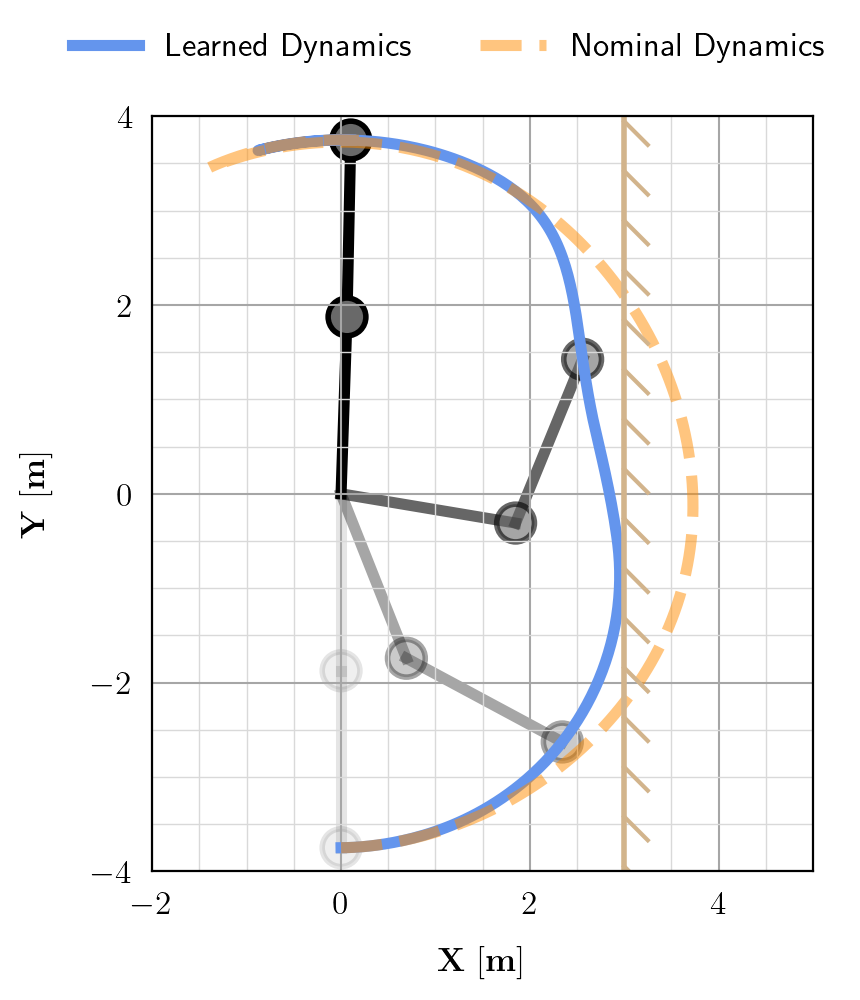}
    \caption{Illustration of the trajectory generated by the learned CBF-QP controller. The solid blue curve represents the end-effector trajectory using the learned dynamics. The dashed orange curve represents the end-effector trajectory using the nominal dynamics. The wall is positioned at $x = 3$. To illustrate the motion of the arm, the two link arm illustrations are drawn such that darker colors correspond to later in time.}
    \label{fig:twolinkarm}
\end{figure}

The two-link arm has the system dynamics
\begin{equation}
    \begin{bmatrix}
        \dot{\mathbf{q}}\\
        \ddot{\mathbf{q}}
    \end{bmatrix} = \begin{bmatrix}
        \dot{\mathbf{q}}\\
        -\textbf{M}^{-1}(\mathbf{q})\textbf{C}(\mathbf{q}, \dot{\mathbf{q}})\dot{\mathbf{q}}
    \end{bmatrix} + \begin{bmatrix}
        \mathbf{0}\\
        \textbf{M}^{-1}(\mathbf{q})
    \end{bmatrix}\boldsymbol\tau,
\end{equation}
where the joint angles are represented by $\mathbf{q}\in\mathbb{R}^2$, the joint velocities by $\dot{\mathbf{q}}\in\mathbb{R}^2$, the joint accelerations by $\ddot{\mathbf{q}}\in\mathbb{R}^2$, and the joint torques by $\boldsymbol\tau\in\mathbb{R}^2$. The inertia matrix is represented by $\mathbf{M}(\mathbf{q})\in\mathbb{R}^{2\times2}$ and the Coriolis matrix is represented by $\mathbf{C}(\mathbf{q}, \dot{\mathbf{q}})\in\mathbb{R}^{2\times2}$. Note that both $\mathbf{M}(\mathbf{q})$ and $\mathbf{C}(\mathbf{q}, \dot{\mathbf{q}})$ are functions of the link masses $m_i$'s and link lengths $\ell_i$'s, for $i = \{1, 2\}$:
\begingroup
\allowdisplaybreaks
\begin{subequations}
\begin{align}
    \mathbf{M}(\mathbf{q}) &= \begin{bmatrix}
        (m_1 + m_2)\ell_1^2 & m_2\ell_1\ell_2\cos(\varphi)\\
        m_2\ell_1\ell_2\cos(\varphi) & m_2\ell_2^2
    \end{bmatrix},\\
    \textbf{C}(\mathbf{q}, \dot{\mathbf{q}}) &= \begin{bmatrix}
            0 & m_2\ell_1\ell_2\dot{q}_2\sin(\varphi)\\
            -m_2\ell_1\ell_2\dot{q}_1\sin(\varphi) & 0
        \end{bmatrix},
\end{align}
\end{subequations}
\endgroup
where $q_1$ represents the angle between the negative $y$ direction and the first link counterclockwise, $q_2$ represents the angle between the negative $y$ direction and the second link counterclockwise, and $\varphi = q_1 - q_2$. The end-effector position can be written as
\begin{equation}
    \begin{bmatrix}
        x_{ee}\\
        y_{ee}
    \end{bmatrix} = \begin{bmatrix}
        \ell_1\sin(q_1) + \ell_2\sin(q_2)\\
        -\ell_1\cos(q_1) - \ell_2\cos(q_2)
    \end{bmatrix}.
\end{equation}
In this example, we assume that the link lengths $\ell_1$ and $\ell_2$ are unknown, and we have access to measurements of the end-effector position. Although the link lengths can be found through inverse kinematics, we use our proposed approach to estimate the system dynamics directly. The system starts from the joint angles $[0, 0]$ and needs to go to $[\pi, \pi]$ while avoiding hitting a wall at $x = 3$. We can write the corresponding HCBF as
\begin{align}
    \widehat{h}(\x) =& -(\widehat\ell_1\cos(q_1)\dot{q}_1 + \widehat\ell_2\cos(q_2)\dot{q}_2) + 3\gamma\nonumber\\
    &- \gamma(\widehat\ell_1\sin(q_1) + \widehat\ell_2\sin(q_2)).
\end{align}

\begin{figure}[t!]
    \centering
    \includegraphics[width=0.49\textwidth]{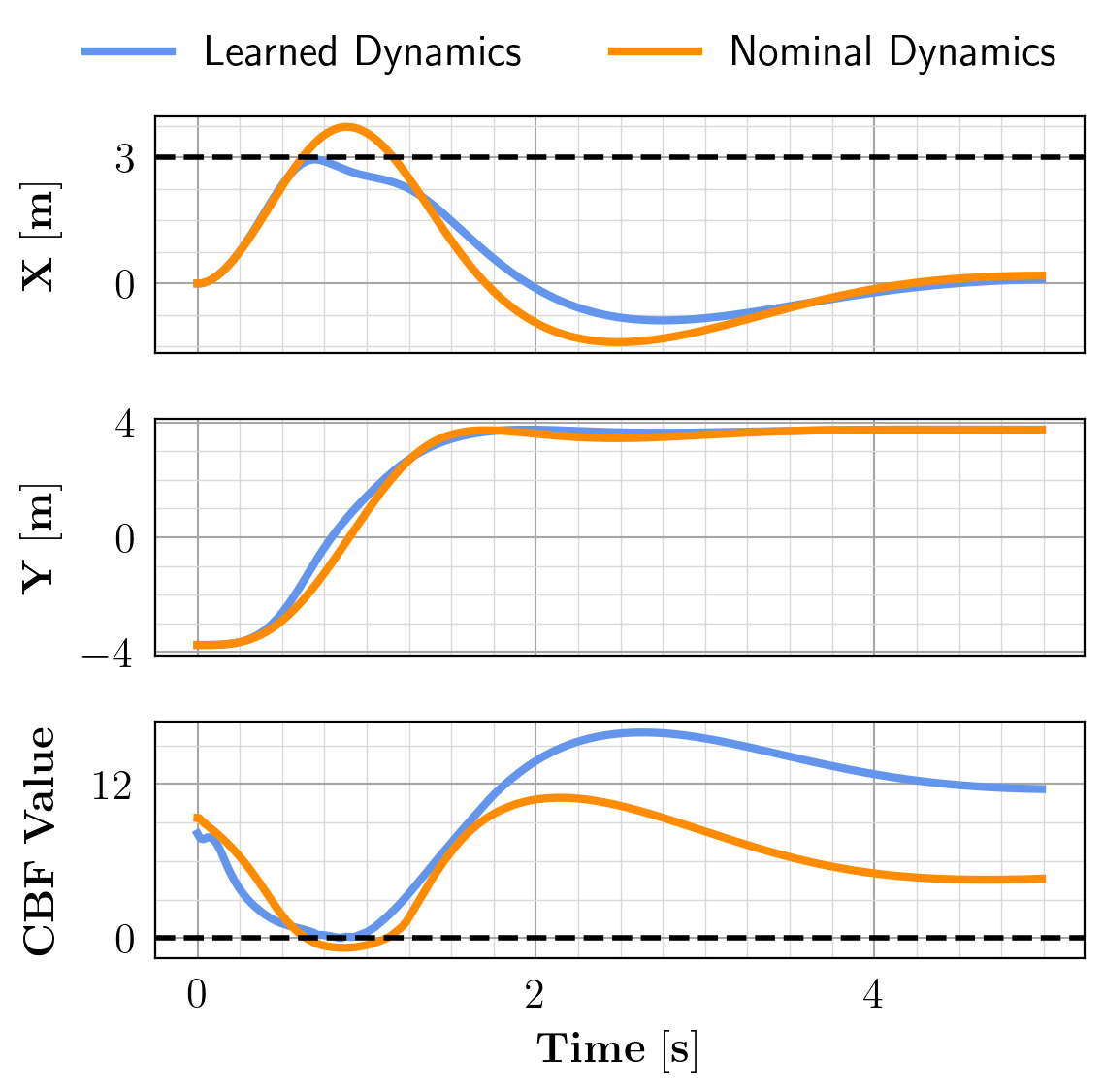}
    \caption{Illustration of the end-effector trajectory and CBF value over time. The dashed line in the uppermost plot represents the position of the wall. The dashed line in the lowermost plot represents the zero-CBF-value line; everything below this line is considered unsafe by the CBF.}
    \label{fig:twolinkarm_data}
\end{figure}

\begin{figure}[t!]
    \centering
    \includegraphics[width=0.49\textwidth]{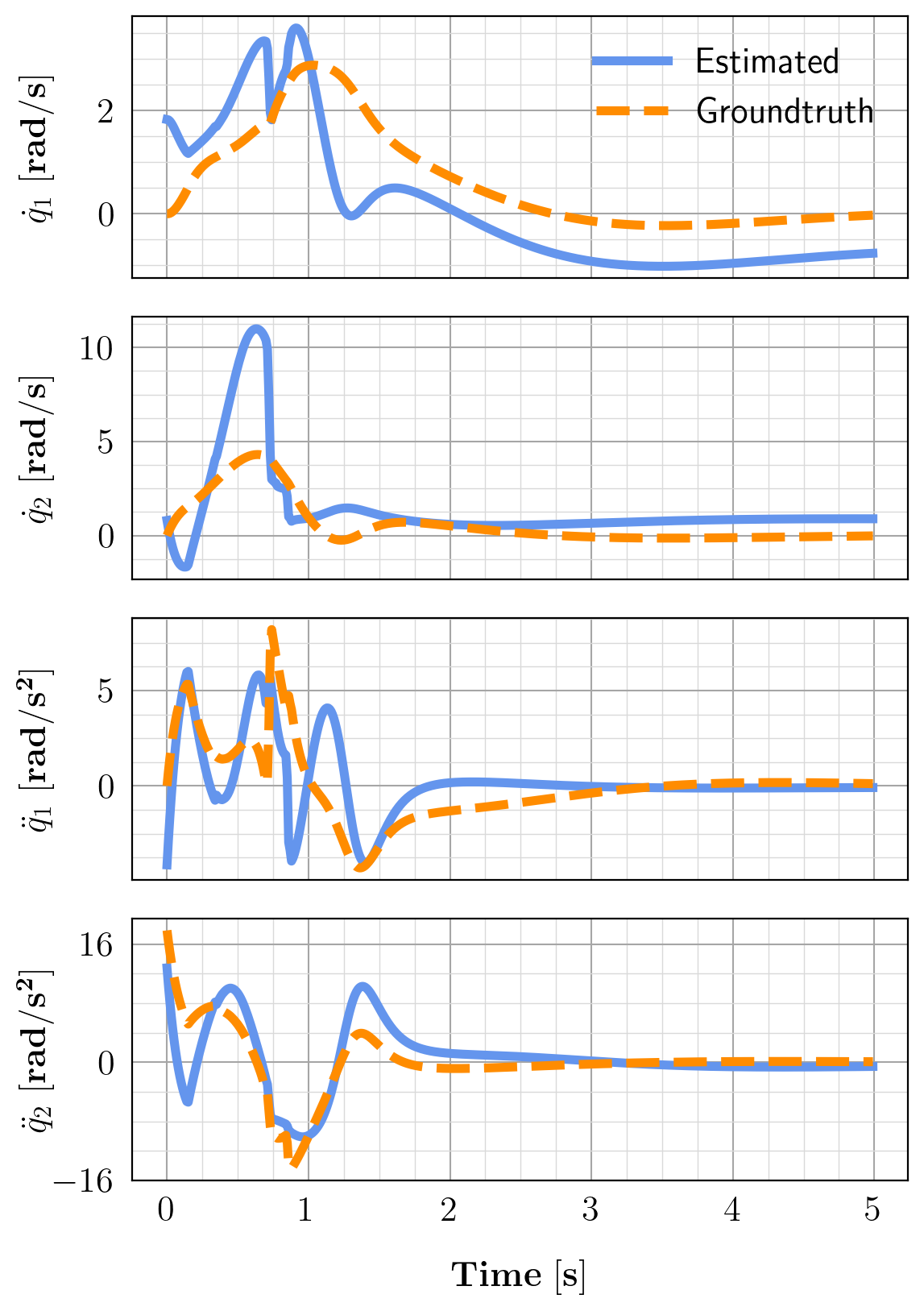}
    \caption{Comparison between the state trajectories of the groundtruth dynamics and the estimated (learned) dynamics for the two-link arm system.}
    \label{fig:twolinkarm_model_est}
\end{figure}

where $\widehat\ell_1 = 1.5$ and $\widehat\ell_2 = 1.5$ are the nominal link lengths and the true link lengths used in the simulation are $\ell_1 = 1.25\widehat\ell_1$ and $\ell_2 = 1.25\widehat\ell_2$. During training, we use a PD controller as the performance controller
\begin{equation}
    \boldsymbol\tau = \begin{bmatrix}
        \mathbf{K}_p(q_1^{\mathrm{des} - q_1}) + \mathbf{K}_d(\dot{q}_1^{\mathrm{des} - \dot{q}_1})\\
        \mathbf{K}_p(q_2^{\mathrm{des} - q_2}) + \mathbf{K}_d(\dot{q}_2^{\mathrm{des} - \dot{q}_2})
    \end{bmatrix},
\end{equation}
with $\mathbf{K}_p = 20.0$ and $\mathbf{K}_d = 15.0$. The loss parameters are chosen as $\lambda_1 = 100.0$ and $\lambda_2 = 0.0$. The learning rate is set to $10^{-5}$. We set 
\begin{equation}
    \mathbf{d}_{+}(\x) = \mathbf{d}_{-}(\x) = 3 - x_{ee}. 
\end{equation}
The class $\mathcal{K}_\infty$ function $\alpha$ is the same as in~\eqref{eq:gamma_func}. The neural networks are trained for 1000 epochs, and the trajectory generated by the learned CBF-QP controller is shown in Fig.~\ref{fig:twolinkarm}. It can be seen that the learned CBF-QP controller can render the system safe while ensuring task completion. 

To study the improvement in robustness attained by learning the system dynamics, we now consider the same training procedure for $\Delta\hat{h}$, but using only the nominal dynamics. In that case, the resulting trajectory is also shown in Fig.~\ref{fig:twolinkarm}, in which we can see the trajectory is unsafe. The end-effector trajectory and the CBF value along the trajectory are shown in Fig.~\ref{fig:twolinkarm_data}. It shows that when using the learned dynamics, as the end-effector position gets closer to the wall, the CBF value goes to zero, and as it leaves the wall, the CBF value increases, which is the expected behavior. When using the nominal dynamics, although the learned CBF can recognize the states are unsafe, the CBF-QP would not be able to generate control actions that pull the system back into the safe set due to having inaccurate system dynamics. The difference between the learned (estimated) dynamics and the groundtruth dynamics is shown in Fig.~\ref{fig:twolinkarm_model_est}. It can be seen that the state trajectories generated by learned dynamics resemble the state trajectories generated by groundtruth dynamics. 
\section{Conclusion}
In this paper, we proposed an algorithmic approach to simultaneously learn a CBF and the system dynamics, starting from an HCBF and nominal dynamics. The CBF is learned using loss functions that enforce the CBF conditions and the CBF constraint. We showed theoretically that our proposed approach could also learn the system dynamics by only using the learned CBF and its time derivative. The effectiveness of our proposed approach is demonstrated using three simulation studies: double integrator target reaching under velocity constraint, unicycle target reaching while avoiding a square obstacle, and two-link arm target reaching while avoiding collision with a wall. In future works, we plan to add a learned performance controller and perform experiments on robotic systems in real life.

\bibliographystyle{plain}        
\bibliography{autosam}           
\end{document}